\documentstyle[psfig,epsf]{mn} 
\def\etal{et al.\ }

\def\msol {\rm{M}$_\odot$} 
\def\mdot {\rm{M}$_\odot$~yr$^{-1}$}

\def\bd{BD+30$^\circ$3639} 
\def\oiii {[O~{\sc iii}]} 
\def\nii {[N~{\sc ii}]} 
\def\sii {[S~{\sc ii}]} 
\def\oiiia {[O~{\sc iii}]5007~\AA} 
\def\oiiib {[O~{\sc iii}]4959~\AA} 
\def\niia {[N~{\sc ii}]6584~\AA} 
\def\niib {[N~{\sc ii}]6548~\AA} 
\def\niic {[N~{\sc ii}]5754~\AA} 
 
\def\siib {[S~{\sc ii}]6731~\AA} 
\def\siir {[S~{\sc ii}]6716~\AA/6731~\AA} 
\def\ha {H$\alpha$} 
\def\hb {H$\beta$} 
\def\kms{\ifmmode{~{\rm km\,s}^{-1}}\else{~km s$^{-1}$}\fi} 
\def\deg{\ifmmode{^{\circ}}\else{$^{\circ}$}\fi} 
\def\vsys{\ifmmode{~v_{\rm sys}}\else{$v_{\rm sys}$}\fi} 
\def\yr{\ifmmode{~{\rm yr}^{-1}}\else{~yr$^{-1}$}\fi} 
\title[Kinematics of BD+30$^\circ$3639] 
{The kinematics of the planetary nebula BD+30$^{\circ}$3639} 
\author[Bryce \& Mellema] 
{Myfanwy Bryce$^1$ and Garrelt Mellema$^2$\\ 
$^1$Department of Physics and Astronomy, The University of Manchester, 
  Oxford Road, Manchester M13 9PL, UK;\\  
  email: mbryce@ast.man.ac.uk\\ 
$^2$Stockholm Observatory, SE-13336 Saltsj{\"o}baden, Sweden;\\  
    email: garrelt@astro.su.se} 
 
\date{Accepted 1999 ---. 
      Received 1998 ---} 
 
\pagerange{\pageref{firstpage}--\pageref{lastpage}} 
\pubyear{1998} 
\begin{document} 
\maketitle 
\label{firstpage} 
\begin{abstract} 
 
In this paper we describe the results of the first optical kinematic 
study of the 
planetary nebula BD+30$^\circ$3639. This system has a central star of the 
Wolf-Rayet type and is believed to be fairly young. Emission line spectra 
were obtained at high spectral and spatial resolution using the Utrecht 
echelle spectrometer at the William Herschel Telescope.  These spectra 
indicate that the main ionized shell of \bd\ appears to be evolving in a more 
complex way than previously thought. 
\end{abstract} 
 
 
 
\section{Introduction}  
The planetary nebula (henceforth PN) BD+30$^\circ$3639 (PN~G064.7+05.0) is a 
compact, young PN with a Wolf-Rayet type central star. Its surface brightness 
is among the highest of all PNe and the object has been intensively studied 
at many wavelengths; from X-rays \cite{ABH96}, optical \cite{HLWB97}, 
infra-red \cite{Latterea95,SZK94}, millimeter \cite{Bachilea91} to radio 
\cite{Bryceea97a}, to name only a few of the many references. 
 
It is believed to be young because of the low effective temperature of the 
star ($\sim 40\,000$~K) and low expansion age of the nebula.  The central 
star is special in that it shows a hydrogen-deficient, Wolf-Rayet type 
spectrum (type [WC9]). The stellar spectrum was analysed by Leuenhagen,
Hamann \& Jeffrey \shortcite{LHJ96} who derived 
a surface temperature ($T_{2/3}$) of 42,000~K, a luminosity, 
$\log_{10}L=4.71$, a mass loss rate of $1.3\times 10^{-5}$ \mdot, and a wind 
velocity of 700 km~s$^{-1}$. $L$ and $\dot{M}$ are distance dependent, and 
these authors assumed a distance of 2.68~kpc to the central star (see
below). As usual with winds 
from WR type stars, BD+30$^\circ$3639's wind is rather massive. 
 
Its distance has been determined using the expansion proper motions. Two
groups used 6~cm {\it VLA}\/ images combined with the optical expansion
velocity from the literature to determine the distance. The reported
distances are $2.8^{+4.7}_{-1.2}$~kpc \cite{Masson89}, $2.680\pm 0.810$~kpc
\cite{Hajianea93}, and $1.5\pm 0.4$~kpc \cite{KawaMass96}. These distances
are substantially larger than the previously derived distances using
statistical methods, which lie mostly in the 600 to 700 pc range
\cite{AckerCat}. Although the proper motion distances should be more reliable
they are somewhat uncomfortable from the point of view of the evolutionary
status of BD+30$^\circ$3639. With a distance of 2.68~kpc, its luminosity
would be around 50,000~$L_\odot$, highest among PNe. Such a high luminosity
would imply a massive ($>1.2$ \msol) central star, which would evolve on very
short time scales ($<100$~years), something which has not been observed. Then
again none of the published stellar evolution models apply directly to
WR-type central stars.
 
BD+30$^\circ$3639 is reported to have some more peculiar properties,
such as a CO expansion velocity substantially higher (50\kms) than
that measured in any other species including H$_2$
\cite{Bachilea91,Shupeea98}. It also shows signs of both C and O-rich
dust, the star itself being C-rich \cite{Watersea98}.
 
Although \bd\ has been extremely well studied, to date no systematic 
kinematic study of the classical optical nebular lines has been done, perhaps 
partly because of the nebula's small size ($6\arcsec\times 4\arcsec$). However, the 
kinematics are an important element in understanding the PN. They can be used 
to help in deprojecting the object, are also essential in converting the 
observed proper motions into distances, and can help in the interpretation of 
the kinematics observed at other wavelengths (such as H$_2$ and CO). In this 
paper we report on the first long slit spectroscopy of BD+30$^\circ$3639. 
 
The lay-out of the paper is as follows: the observations 
and data analysis are described in Sect.~2, 
in Sect.~3 we discuss and interpret the results, and we 
sum up our conclusions in Sect.~4. 
 
\section{Observations and data analysis} 
High dispersion, spatially resolved spectra were obtained from \bd\ using the 
William Herschel Telescope (WHT) on La Palma combined with the Utrecht 
echelle spectrometer (UES) and the Tek1 CCD detector.  The observations were 
made on 1994 August 20 and 21.  UES was used with the 79~grooves~mm$^{-1}$ 
grating and the maximum slit length of 15\arcsec, long enough to cover the 
main shell of \bd. However, the wide inter-order separation which permits use 
of a relatively long slit also means that full order coverage is not possible 
with the $1024 \times 1024$ pixel Tek1 CCD.  Therefore, each slit position 
was observed at two grating positions, covering either end of the $31^{\rm 
st}-45^{\rm th}$ orders (4740 \AA\ -- 7415 \AA) which include most of the 
bright, kinematically and diagnostically useful optical emission lines.  A 
slit width of 252~$\mu$m ($\equiv 1.12\arcsec$) was used; this is matched to 
two detector pixels (each $24~\mu\mbox{m} \times 24~\mu\mbox{m}$), giving a 
velocity resolution of 6.12\kms.  In the spatial direction (with the 
derotator), the scale at the detector is 14.96\arcsec\ mm$^{-1}$, so that 1 
pixel $\equiv 0.36\arcsec$.  The `seeing' remained consistent at $\sim 
1\arcsec$ throughout the observing period. 
 
\begin{figure} 
\centering 
\epsfclipon 
\mbox{\epsfxsize=3.2in\epsfbox[95 23 585 400]{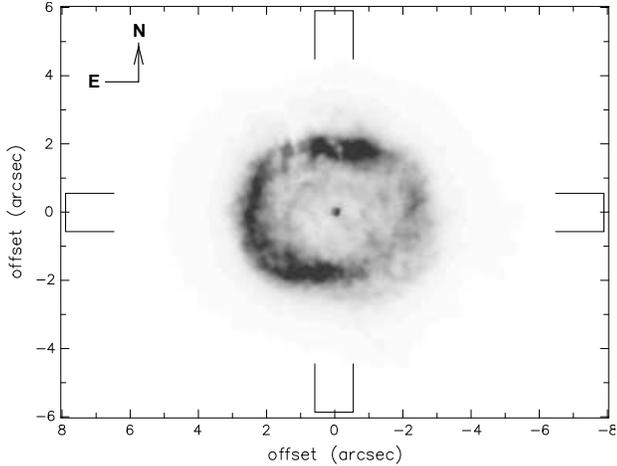}} 
\caption{The HST narrow band \ha\ image of \bd\ (Harrington et
al.~1997) with the EW and NS slit orientations indicated.} 
\label{f_image+slits} 
\end{figure} 
 
The main ionized structure of this PN is 
remarkably rectangular in appearance with its major and minor axes aligned 
approximately along the EW and NS directions respectively
(Fig.~\ref{f_image+slits}). 
Data were obtained from several slit positions aligned NS and EW across the 
nebula. A Thorium-Argon 
wavelength calibration spectrum was obtained at each slit and grating 
position.  Flat-field frames were obtained using a quartz lamp and a wide 
slit ($\equiv 5\arcsec$) exposure of the flux standard star SP~1942+261 was 
also obtained at each grating position. 
  
When these observations were obtained, the A\&G system on the UES was
not fully operational and it became apparent after the data reduction
process that we could only be absolutely certain of the spatial
positions of the three runs which were clearly centred on the central
star (Table~\ref{t_data}).

\begin{table}
\caption{Observational Data}
\label{t_data}
\begin{tabular}{@{}llll}
 & run 1 & run 2 & run 3 \\ \hline
slit orientation & E-W & N-S & N-S \\
& & & \\
exp time (s)& 600 & 60 & 600  \\
& & & \\
emission lines & *\niia  &  \niia  & \sii 6716+6731\AA \\
(* - saturated) &  *\ha & \ha  & *\ha\\ 
& He~{\sc I}~5876\AA\ & He~{\sc I}~5876\AA\ & \niib\\
&  \niic &  \niic & [O~{\sc I}]6364\AA\\
& [N~{\sc I}]~5198+5200\AA\ &[N~{\sc I}]~5198+5200\AA\ & He~{\sc I}~5876\AA\ \\
 & \oiiia &\oiiia & \oiiib\\
 &  \hb &  \hb & \hb\\ \hline
\end{tabular}
\end{table}

The advantage of 
obtaining data from several prominent emission lines in one exposure was 
largely cancelled out by the loss of data due to positional uncertainty and 
the much longer timescale required for the non-standard reduction 
process. These factors should be considered if future observations of this 
type are undertaken. 
 
\subsection{Data reduction} 
The initial processing was done using standard IRAF routines to debias the 
data and calibration frames and to combine multiple exposures using standard 
cosmic ray rejection algorithms.  No dark frames were obtained since the Tek 
CCD has a very low dark current. 
The remainder of the processing was done using STARLINK routines.  The data
were flat fielded in the usual way before the individual
position-velocity (pv) arrays were extracted and wavelength calibrated.
This part of the reduction process was non-standard and is 
described in Appendix A.
Although some 
scattering was apparent in the inter-order separations around the bright 
emission lines, this was not removed since the 2-d nature of the spectra made 
it difficult to generate an accurate model for subtraction.  The scattering 
is at a relatively low level and does not affect the kinematical 
interpretation of the spectra significantly. 
Finally the spectra were flux calibrated and the spectrum of the
central star was removed from the pv arrays (Appendix B).   
 
\begin{figure*} 
\centering 
\epsfclipon 
\mbox{\epsfxsize=6.3in\epsfbox[9 36 525 777]{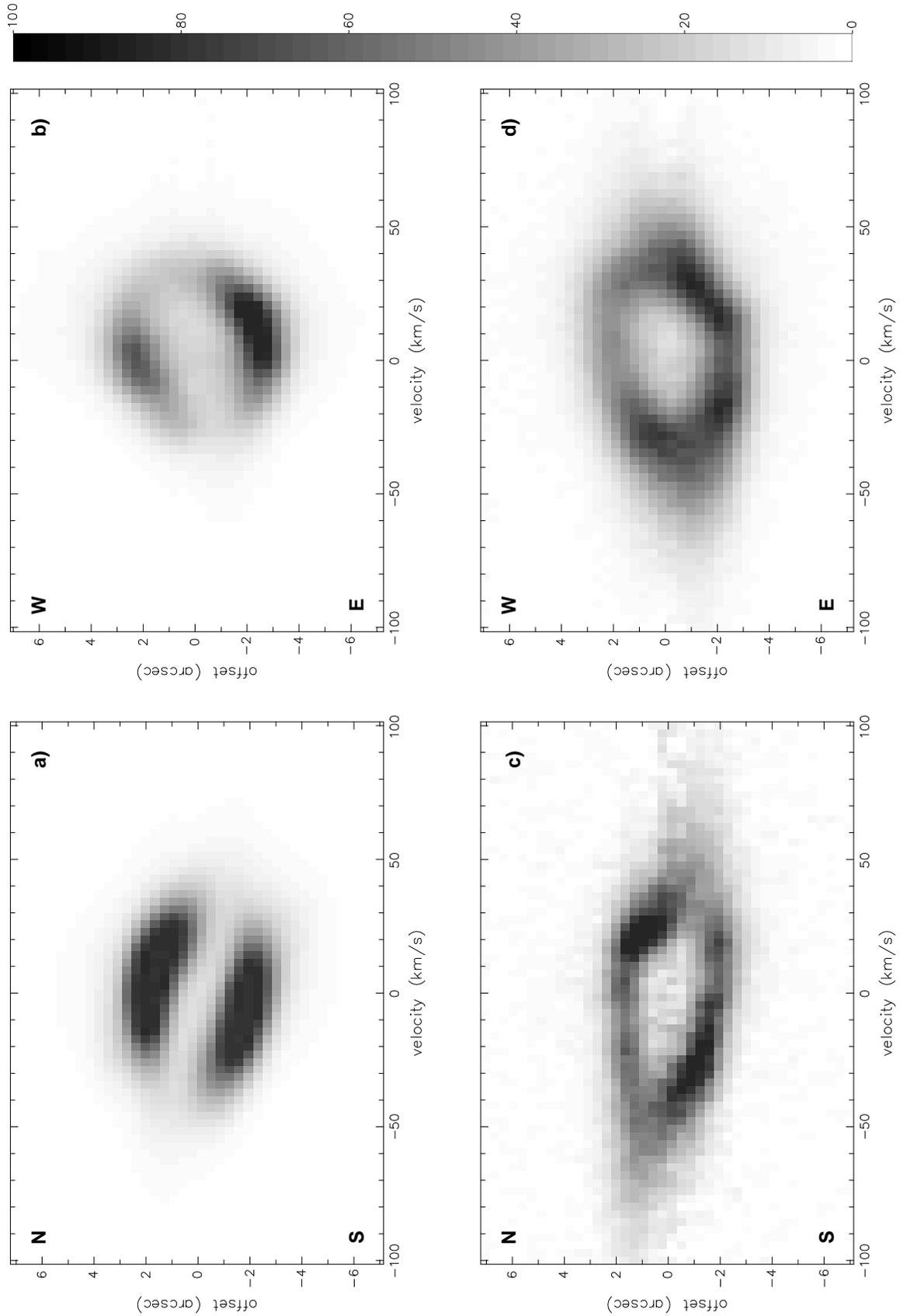}} 
\caption{The two-dimensional position-velocity arrays of the
(a)\niib\ (0--5000), (b)\niic\ (0--100), (c)\oiiib\ 
(0--15) and (d)\oiiia\ (0--36) emission obtained from runs~1 and~3.  The 
scale bar  corresponds to the ranges indicated, which are in 
units of mJy per pixel.  } 
\label{figs_fig2} 
\end{figure*} 
 
\subsection{The main kinematical features} 
Examples of the spatially resolved [O~{\sc iii}] and [N~{\sc ii}] emission 
line profiles obtained after the reduction process from the NS and EW slits 
are given in Fig.~\ref{figs_fig2}. The wavelength axis on each image has 
been converted to radial velocity offset from the systemic velocity $\vsys$ 
which was chosen to correspond to a heliocentric velocity of $-32\kms$.  The 
spatial scale on each image has been converted to an offset in arcseconds 
from the position of the central star. In each case, the stellar spectrum has 
been subtracted from the data frame (Appendix B). 
As noted previously, it became clear when the data was analysed that
the absolute slit positions of several of our observations could not
be determined.  Although we obtained unsaturated \niic\ data at both
slit orientations, this line is intrinsically faint and so we prefer
to show the much brighter  \niib\ pv array obtained in run~3 
in Fig.~\ref{figs_fig2}a) rather than the \niic\ pv array obtained
in run~2. Although the relative brightnesses of these two
emission lines depends on the electron density of the emitting gas,
there is no significant difference between the shapes of the two pv
arrays. Similarly, we obtained \oiiia\ data at both E-W
and N-W slit positions but again this line is intrinsically weak and the short 
exposure time for the N-S position (run 2) gave a very noisy
profile.  Hence we prefer to show the \oiiib\ pv array obtained with a 
longer exposure time (run 3) in Fig.~\ref{figs_fig2}c).  
There is no problem in the direct
comparison of these two lines as the \oiiia\ and \oiiib\ emission
lines occur in the ratio of 3:1.
 
The main \nii\ shell of emission is more spatially extended but less 
spectrally extended than the main \oiii\ shell.  The velocity ellipses in 
both high and low ioniziations and from both spatial directions are tilted 
with respect to the spectral axis and indeed the \nii\ velocity ellipses 
appear to be almost open ended, with \oiii\ emission emerging from these 
gaps. 
 
The NS and EW slit orientations were chosen because these are the projected
minor and major axes of the PN (see Fig.~\ref{f_image+slits}). For a
cylindrically symmetric nebula, these projected axes coincide with the
direction of the three--dimensional (equatorial) symmetry plane and the
(rotational) symmetry axis.  However, the spatio-kinematic line emission maps
shown in Fig.~\ref{figs_fig2} indicate that this may not be the case.  The
two slit positions show that the blue shifted component of the bright \nii\
emission from the NE part of the nebula is absent and similarly the red
shifted component is absent from the SE part of the nebula.  However, high
speed blue and red shifted components appear in the \oiii\ emission from the
NE and SW regions respectively. Similar high speed components were also found
in the \nii\ pv arrays although these components are much fainter than the
main \nii\ features shown in Fig.~\ref{figs_fig2}.  Since the spatial offsets
of the fastest moving components are close to the centre of the nebula, this
implies a low inclination angle of a prolate nebula. The implications of this
are discussed further in Sect.~3.
  
\begin{figure} 
\centering 
\epsfclipon 
\mbox{\epsfxsize=3in\epsfbox[90 72 477 577]{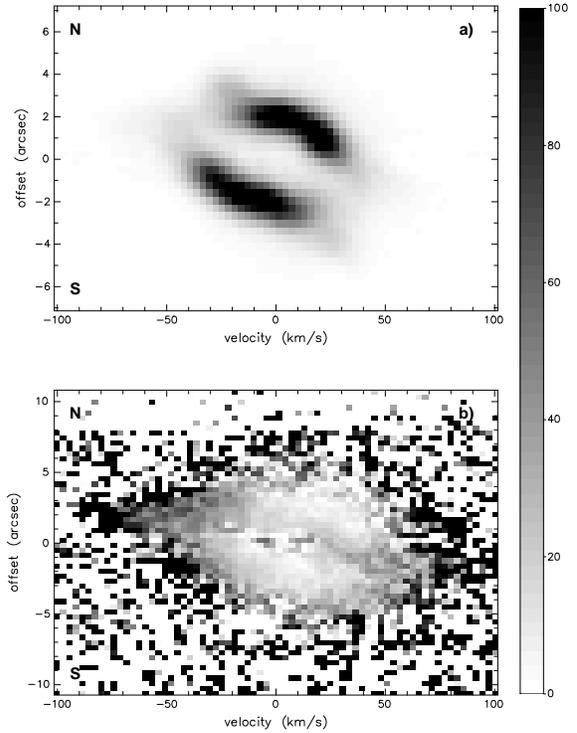}} 
\caption{(a) Position velocity array of the \siib\ emission, scaled 
between 0 and 200 mJy/pixel.  (b) \siir\ ratio map scaled between 0.4 and 
0.8. The scale bar corresponds to these ranges. 
} 
\label{figs_fig3} 
\end{figure} 
An \siir\ ratio map was made from the data from the NS slit and is shown in
Fig.~\ref{figs_fig3}, together with the \siib\ position-velocity array.  Note
that the \siib\ pv array shows the same open ended velocity ellipse structure
as that seen in the \niib\ pv array shown in Fig.~\ref{figs_fig2}a).  The
\siir\ intensity ratio (at a temperature of $10\,000$ K), varies from $\sim
1.45$ to $\sim 0.45$ over an electron density range of $\sim 10 - 10^5 \rm{
cm}^{-3}$ \cite{ost89}.  The ratio map shows that there is a real variation
in electron density around the velocity ellipse, with the apparently
brightest parts having electron densities $> 2 \times 10^4 \rm{~cm}^{-3}$,
dropping to $3-7 \times 10^3 \rm{~cm}^{-3}$ at the fainter ends.  The very
faint, higher speed, blue-shifted component apparently emanating from the
northern opening displays an electron density $<1000 \rm{~cm}^{-3}$.
 
\section{Discussion} 
\subsection{Nebular morphology and deprojection} 
The appearance of the line shapes as presented in the previous section have
some interesting implications for the interpretation of the morphology of the
nebula. We placed the slits along the projected minor and major axes of the
nebula, which in the case of an inclined, cylindrically symmetric nebula
would result in a skewed velocity ellipse along the major axis and a
symmetric velocity ellipse along the minor axis. Our results show this not to
be the case. Considering the {\it HST}\/ and {\it VLA/MERLIN}\/ images one
may question whether BD+30$^\circ$3639 is cylindrically symmetric since it
looks more like a rounded off rectangle, something which appears to be
inconsistent with any kind of projection of a cylindrically symmetric shell
\cite{Bryceea97a,HLWB97}.  However, even with this boxy shape one would
expect the line shapes to contain some information about the orientation of
the bubble. Combining the NS and EW slits shows that the highest redshifts
are found in the SW and the highest blueshifts in the NE part of the
nebula. The most straightforward interpretation is then that the (rotational)
symmetry axis of the nebula runs more or less diagonally across it roughly
from NE to SW, with the NE side approaching us.
 
The brightness distribution in the line shapes, as well as the electron
densities derived from the \siir\ intensity ratio show that the fastest parts
of the PN also have the lowest densities, which is consistent with the
picture of a prolate nebula with faster, low density polar regions. The
morphology of \bd\ combined with shape of the lines shows that we are seeing
the nebula almost pole-on. Similar line shapes lead to this conclusion in the
case of the Ring Nebula NGC~6720 \cite{BryBaMae94}. Further evidence comes
from a comparison with the model line shapes based on hydrodynamic
simulations presented in Chapter~5 of Mellema \shortcite{Mellema93} and the
simple ellipsoidal model based on the distribution of the radio flux
\cite{Masson89}, which gave a best fit inclination of $18^\circ$ (but note
that he assumed the rotational symmetry axis to run EW). We do not think that
the low spatial resolution of the data justifies more detailed kinematic
modelling than this.
 
This `deprojection' is supported by some other observations. As was pointed
out in Bryce \etal \shortcite{Bryceea97a} the smooth extinction gradient
across the nebula (as derived from comparing high resolution optical and
radio emission, see also Harrington \etal \shortcite{HLWB97}), can also be
interpreted as being due to inclination. Assuming that most of the dust is
located in a halo surrounding the entire optical nebula, or in an equatorial
torus, and that the near side of the nebula is the NE, one would expect
higher dust extinction in the SW, as is observed. The high extinction clump
seen in the north-eastern part could then be lying along the rotational
symmetry axis and be the feature in \bd\ which lies closest to us.
 
Graham \etal \shortcite{Grahamea93} reported the detection of an H$_2$ blob
NE of the main optical nebula which was confirmed by Shupe \etal
\shortcite{Shupeea98}. There is some indication of a structure at that
position in the {\it HST}\/ \sii/\ha\ ratio map presented by Harrington et
al.~(1997). Sahai \& Trauger \shortcite{SahaiTr98} present the {\it HST} data
in a way which shows the faint small scale structures in the halo of
\bd. Here one sees radial structures in the SW part of the halo and perhaps a
smoother structure in the NE part. Although none of these observations
individually could support the claim that we are seeing a prolate nebula
almost pole-on with the rotational symmetry axis running NE-SW, combined this
seems to be the most likely interpretation.
 
\begin{figure} 
\centering 
\epsfclipon 
\mbox{\epsfxsize=2.5in\epsfbox[0 0 400 350]{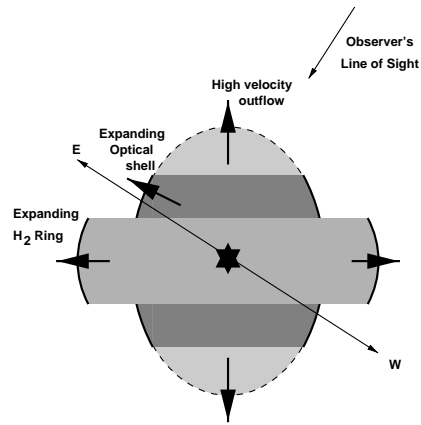}} 
\caption{A sketch illustrating one possible scenario for the
kinematical structure of \bd.  Moving eastwards from the central star, 
the observer sees firstly the fast-moving, low density, optical
emission along the approaching polar axis of the PN, then the lower
ionization, bright optical emission from the main nebular shell
expanding almost tangentially to the line of sight and
finally the H$_2$ emission from an outer ring of gas with a larger
component of velocity along the line of sight.
} 
\label{figs_fig4} 
\end{figure}

Comparing our new optical kinematic data to the H$_2$ 1-0 $S(0)$ data in
Shupe \etal \shortcite{Shupeea98}, we do encounter a problem. Basically these
authors find the opposite kinematic behaviour to that exhibited in our data:
the H$_2$ emission to the east of the nebula is red--shifted by up to 30\kms\
with a faint component reaching 60\kms\ and that to the west is
blue--shifted, again by up to 30\kms\ from the systemic velocity. However,
our faint, high speed optical components are blue-shifted towards the east
and red-shifted towards the west (the ionized gas at the eastern and western
edges of the main optical shell is moving almost tangentially to the
line-of-sight). We checked our slit orientation carefully and communicated
with the authors of the other paper and the effect seems to be genuine. The
explanation could be that the H$_2$ lies in an expanding flattened structure
around the prolate nebula, in which case it would be possible to see both
positive and negative velocities along the same position angle, due to the
effects of superposition along the line of sight. Such a scheme is
illustrated in Fig.~\ref{figs_fig4}.  The H$_2$ map presented in Shupe \etal
\shortcite{Shupeea98} supports this interpretation since it shows a ring-like
distribution around the ionized nebula. If this is the case the expansion
velocity of parts of this ring would have to be quite high: the projected
expansion velocity is about 30\kms, which with the estimated inclination
angle leads to a physical expansion velocity which is 2 to 3 times
higher. Unfortunately, the resolution of the H$_2$ data is only 50\kms, so it
is difficult to compare the molecular and ionized kinematics one to one.

\subsection{Ionization stratification} 
Already in narrow band images it can be seen that there is a difference 
between the low (\nii) and high (\oiii) ionization regions of 
the nebula \cite{HLWB97}: the high ionization region is smaller than the low 
ionization region. The kinematic data show an even clearer difference. The 
high ionization lines are spatially less extended, reach much higher 
velocities and show a closed shell morphology, whereas the low ionization 
regions are indicative of an open ended shell (see Fig.~2). Combined with 
the deprojection suggested in the previous section this means that the 
prolate nebula has fast lower density polar caps which are faint in the low 
ionization lines, but relatively bright in the high ionization lines. 
 
Masson \shortcite{Masson89} suggested that \bd\ might be like
NGC~7027, but seen pole-on. 
This was based on his simple ellipsoidal shell model. Because of the quite 
severe dust extinction in NGC~7027, it is difficult to compare optical narrow 
band images of the two nebulae. There is however a nebula which displays 
exactly the behaviour described in the previous paragraph, namely NGC~40. 
This nebula has an ellipsoidal shape with faint polar caps as seen in low 
ionization lines, whereas the high ionization lines show a closed shell. 
Another similarity between \bd\ and NGC~40 is that both have a WR-type 
central star, type [WC8] and [WC9] respectively \cite{LHJ96}.
 
When discussing the image and velocity data of NGC~40, Meaburn \etal
\shortcite{Meabea96} suggested that the shells seen in [N~{\sc II}] and
[O~{\sc III}] might have a different origin, the first one swept up by an
ionization front, and the second by the fast wind from the WR-type central
star. Nebulae of this type were found in the numerical models of Mellema
\shortcite{Mellema95} and Mellema~\shortcite{Mellema97}.  One expects such a
configuration to occur early in the nebula's evolution, which is consistent
with \bd. The models show that one clear difference between the ionization
and wind driven shells is that the first does not show much velocity
variation as a function of latitude, whereas the latter does. In the case of
NGC~40 the [O~{\sc III}] data were not available to test this hypothesis. For
\bd\ we clearly observe higher velocities in the [O~{\sc III}] lines,
supporting such a scheme. However, the size difference between the nebulae in
the [N~{\sc II}] and [O~{\sc III}] images is rather marginal. As in the case
of NGC~40 this means that the two shell scenario remains speculative.
 
\subsection{Expansion velocities} 
\bd\ is one of the radio brightest PNe and its distance has been measured by
Kawamura \& Masson (1996, $900\pm 200$ pc) and Hajian \etal (1993, $2680 \pm
810$ pc), both using the observed proper motions of the expanding shell of
ionized gas.  In both cases the distance was calculated by combining the
tangential motion derived from observations obtained at different epochs with
the radial expansion velocity derived from optical emission line spectra.
Both papers obtained an angular expansion rate of $\sim 1.3$ mas \yr\ for the
outer edge of the minor axis, and both papers adopted an expansion velocity
of 22\kms. However, Kawamura \& Masson (1996) then included a correction for
a constant flux assumption which explains the discrepancy between the final
distance values.  An expansion velocity of 22\kms\ was originally adopted by
Masson (1989), in an earlier attempt to measure the distance to \bd, who made
the point that good, spatially resolved kinematical data was not available at
that time.
 
\begin{figure} 
\centering 
\epsfclipon 
\mbox{\epsfxsize=3in\epsfbox[30 26 313 602]{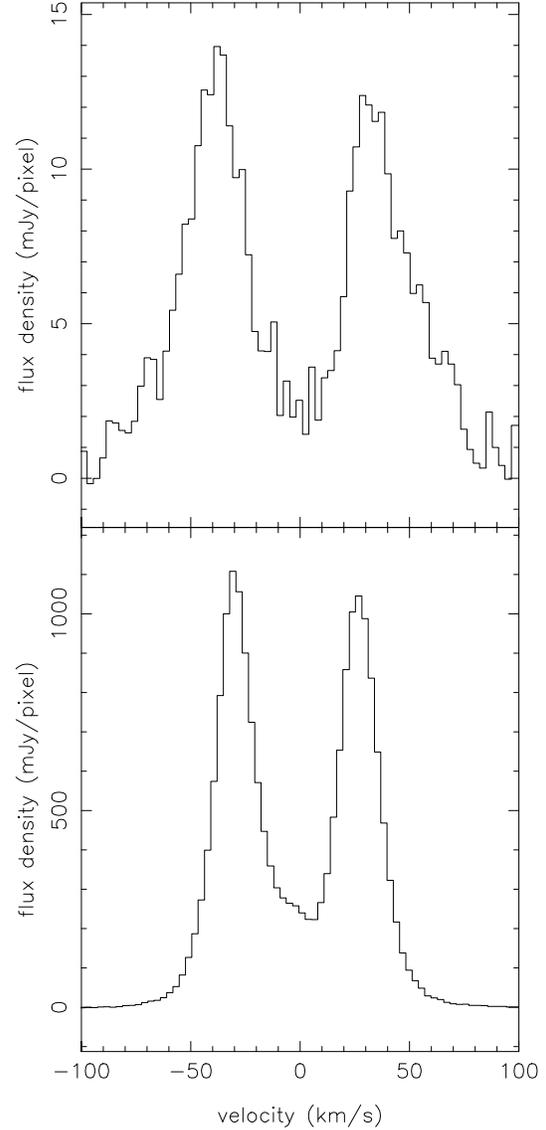}} 
\caption{Spectra obtained by adding together the two central 
crossections from the NS slit for the \oiiib\ (upper) and 
\niib\ (lower) emission lines. 
} 
\label{figs_fig5} 
\end{figure} 
 
Our new data provide much more accurate expansion velocities.  In Fig.~5, 
spectra from the two central crossections of the position-velocity arrays 
shown in Fig.~\ref{figs_fig2}(a) and (c) are presented. In each case the 
spectrum was fitted with Gaussian profiles and values for the radial 
expansion velocity observed at the centre of the nebula were derived from the 
centres of the two Gaussian components corresponding to the two main emission 
components.  This gave values of $28 \pm 1 \kms$ and $35.5 \pm 1 \kms$ for 
the \nii\ and \oiii\ profiles respectively.  The value for the \oiii\ 
expansion is considerably higher than the value of 23\kms\ obtained by 
Sabbadin (1984). 
The explanation for this discrepancy probably lies in the 
fact that if such a measurement is made using lower spatial resolution (i.e.\ 
a wider slit or binning along the slit), 
the brighter, but radially 
slower moving components dominate the spectrum. Aller \& Hyung (1995)
obtained a value of $\sim 26.8 \kms$ using a $1.2\arcsec \times 4
\arcsec$ slit. In a simple proper motion 
measurement, the calculated distance to the object scales directly with 
expansion velocity, indicating that the distance measurements of Kawamura \& 
Masson (1996) and Hajian et al.~(1993) may have to be revised upwards. This 
would make the luminosity problem described in Section 1 even worse. 

However, a serious problem for these types of distance calculation is that 
they are also dependent on a model which relates the radial expansion 
velocity to the observed tangential proper motions. If  we are observing a 
prolate nebula almost pole-on, the discrepancy between the radial polar 
velocity and tangential equatorial proper motions is at its largest. This 
effectively means that the distances derived become upper limits. If the 
outer edge of the PN is actually an expanding ionization front (as suggested 
in Section 3.2), the relation between the radial gas velocity and the 
tangential proper motions becomes even more uncertain. 
 
\subsection{Evidence for a high-speed outflow} 
\begin{figure} 
\centering 
\epsfclipon 
\mbox{\epsfxsize=3in\epsfbox[20 240 230 690]{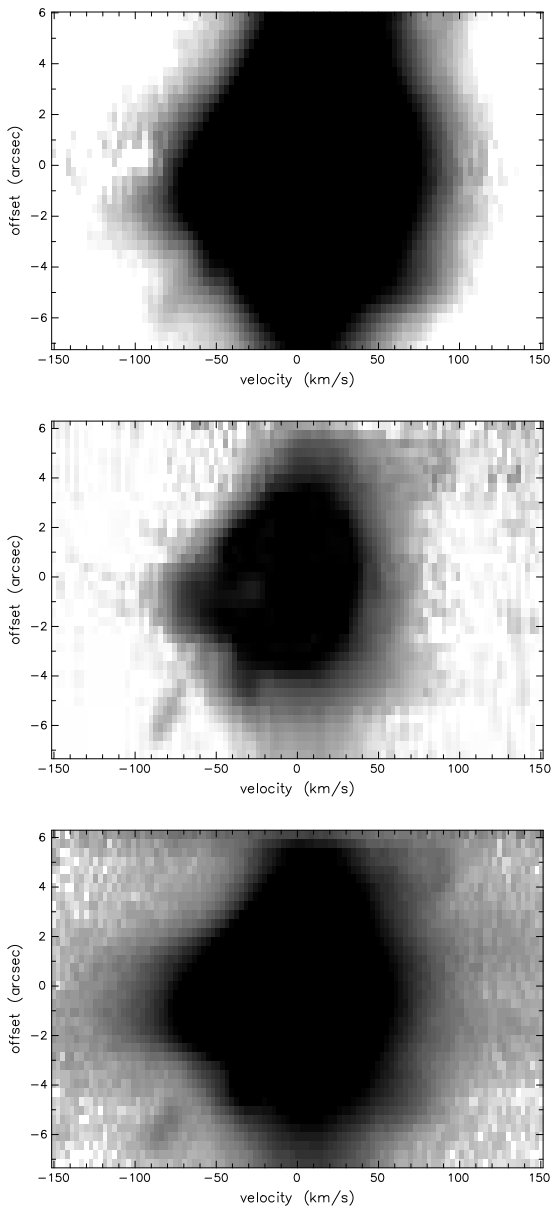}} 
\caption{a) The p-v array of the \niia\ emission from the EW slit (run
1), showing a faint, accelerating flow to the east of the main nebular
shell.  b) The p-v array of the \siib\ emission obtained from an EW
slit position offset slightly north of the central star; the
accelerating flow is more prominent. c) The p-v array of the \niib\
emission from the same slit position as b).  }
\label{figs_fig6} 
\end{figure} 
 
Harrington \etal\ \shortcite{HLWB97}
presented several emission line ratio maps obtained 
from their WFPC2 images of \bd. As mentioned above they discovered a compact 
`east blob', particularly prominent in their \sii/\ha\ ratio map which is 
about twice as far from the centre of the nebula as the eastern limb of the 
main shell and coincides with the region of bright H$_2$ emission 
discovered by Graham \etal~(1993).  

Our data contains an interesting feature which could be associated with this
`east blob'. The \niia\ emission line from the EW slit is shown in
Fig.~\ref{figs_fig6}a, on a logarithmic scale chosen to show up the very
faintest emission surrounding the main, bright, nebular shell
(cf.~Fig.~\ref{figs_fig2}b).  The bright emission in this line from the main
nebular shell saturated the CCD chip, but a faint feature extending towards
the bottom left corner can be discerned.  This feature lies at about
5--6\arcsec\ east of the central star and has a velocity ranging from $\sim
-60$ to $-90 \kms$, much larger than the velocities exhibited by the main
nebular shell.  Also shown are data from another EW slit position, which was
slightly offset (probably north) from the central star.  The fast-moving
feature appears very clearly in the \siib\ emission line
(Fig.~\ref{figs_fig6}b) and also, though less prominently, in the \niib\
emission line (Fig.~\ref{figs_fig6}c).  The intensity observed
in the fast moving feature expressed as a fraction of the intensity
in the brightest parts of the main shell was found to be approximately
5 times higher in \sii\ than in \niib\ which is indicative of shock
excitation. The \siir\ in the accelerating flow indicates an electron density
$<850 \rm{~cm}^{-3}$.
 
Fast, low ionization features are often found outside the main shells of more 
evolved PNe (e.g. FLIERS -- Balick \etal 1993, BRETS -- L{\'o}pez,
Meaburn \&\ Palmer 1993) and 
recently very high speed features were observed around the young PN MyCn~18 
\cite{Bryceea97b}.  In most cases, such knots appear in pairs, on opposite 
sides of the central star, and in cases where more than one pair is observed 
there is often a point symmetry in the pattern.  The feature in \bd\ is only 
clearly seen on one side of the central star.  Further observations 
are needed to establish the properties of this high velocity feature
and to search for a possible counter-feature on the opposite side of
the nebula. 

In Bachiller et al.~(1991;1992) it was noted that the CO outflow had a higher
velocity than the optical outflow. Our data show this to be no longer
true. The velocity of the feature discussed above reaches 90\kms, and even in
the main nebula these kinds of velocities are reached (see for example
Fig.~2c). As noted above, the previously reported lower optical expansion
velocities were based on measurements of the brighter parts of the
nebula. The high velocity optical emission does seem to be strongly localised
in the NE and SW. Bachiller et al.~(1992) suggested two explanations for the
observed CO spectrum: an expanding shell or two discrete bipolar blobs. They
rejected the latter interpretation, but in view of our new results this
should perhaps be reconsidered.
 
\section{Conclusions} 
New, spatially resolved, high spectral resolution, position-velocity maps
have been obtained in both high and low ionization emission lines from the
young PN \bd.  The low ionization data suggest a prolate, nearly axially
symmetric, open-ended shape for the main nebular shell, with the rotational
symmetry axis at about PA 30\deg -- 60\deg\ in the plane of the sky and an
inclination of about 20\deg. The higher ionization material shows a closed
shell structure of a somewhat smaller size. This structure may be similar to
that seen in NGC~40. The high inclination makes it difficult to make the
connection with the previously published image data, which show a remarkably
rectangular shape for the main shell in both optical \cite{HLWB97} and radio
continuum emission \cite{Bryceea97a}.

The velocity signature for the ionized material seems to be exactly opposite
to that of the molecular material \cite{Shupeea98}: the eastern side is
mostly blue-shifted in the ionized lines and red-shifted in the molecular
lines. This behaviour can be explained if the molecular material sits in a
flattened structure around the prolate nebula.

New estimates for the expansion velocity determined along the line of sight 
to the central star have been obtained, $28 \pm 1 \kms$ and $35.5 \pm 1 \kms$ 
for the \nii\ and \oiii\ profiles respectively. These are higher velocities 
than previously supposed, probably because previous observations had a lower 
spatial resolution and therefore incorporated bright emission from the shell 
limbs which has a lower radial velocity. These new values can be of use in 
attempts to obtain the distance to \bd\ using proper motion measurements 
(Kawamura \& Masson 1996; Hajian et al.~1993). However, it should be kept in 
mind that the radial expansion velocity must be related to the tangential 
proper motions via a model of the nebular structure and as noted above, our 
results indicate that a simple geometry may not be appropriate. 
 
A fast moving ($-90 \kms$), faint feature was detected to the east of the
main nebular shell.  This feature may coincide with structure in the halo of
\bd\ detected by Harrington \etal~\shortcite{HLWB97} in WFPC2 images and by
Graham \etal~\shortcite{Grahamea93} from H$_2$ observations. This feature and
the high velocity parts of the main shell have velocities close to twice as
high as the reported CO velocity of $50 \kms$ \cite{Bachilea91}.
 
\section*{Acknowledgments} 
The WHT is operated on the island of La Palma by the Isaac Newton Group in
the Spanish Observatorio del Roque de los Muchachos of the Instituto de
Astrofisica de Canarias.  MB is in receipt of a Manchester University
Research Fellowship. GM acknowledges support from the Swedish Natural Science
Research Council (NFR).  We would like to thank Dr Jeremy Walsh (STECF) for
useful discussions on obtaining electron densities from 
the \siir\ ratio and for providing  us with his
program to calculate electron density at a given temperature for a given
value of the ratio.

\appendix
\section{extraction and calibration of indvidual echelle
orders} Neither STARLINK nor IRAF contain specific packages for reducing
cross-dispersed echelle spectra in the case where the 2-d structure of the
emission line profiles are to be preserved.  This is because, spatially
extended, cross-dispersed spectra are usually obtained for point-like sources
and the extended slit thus permits accurate sky subtraction, a process which
is folded into the standard reduction packages which then produce a 1-d
spectrum of the target.  The first problem with cross-dispersed echelle data
is to straighten the orders. Each order appears as an approximately vertical
stripe in the raw CCD frame but each `stripe' is at a slight angle to the
next one. It is desirable to rebin the data such that the `stripes' become
exactly parallel.  The largest angle between an order and the y-axis was
0.87\deg, equivalent to a shift of 15.5 pixels along the x-axis from the
first crossection to the last.  The amount of rotation required to straighten
the orders was found to be too small for standard rotation routines which did
not handle the rebinning of data sufficiently accurately.  The orders were
straightened using routines which are usually used to SCRUNCH (wavelength
calibrate) data.  We used the continuum spectrum from the standard star as a
pseudo-calibration arc.  This continuum spectrum traces out a set of
approximately vertical lines.  Each `arc line' was identified with the
horizontal position (`rest wavelength') of the continuum spectrum in that
order in the vertically central crossection of the CCD frame, using the ARC2D
routine from the TWODSPEC package in STARLINK.  A dispersion calibration file
was then produced.  This file was used to `SCRUNCH' the other data files,
resulting in a set of CCD frames containing parallel echelle orders.  The
CCDEDIT and TRANNDF routines from the CCDPACK package were used to extract
and rotate (through 90\deg ) each order of interest such that wavelength
increases from left to right, resulting in a set of pv arrays, each with the
central star at the same spatial position.

The PN data frames and standard star frames for each order were then 
wavelength calibrated, following the normal wavelength calibration procedure 
for longslit spectra, using the Th-Ar calibration spectra and the routines 
ARC2D and ISCRUNCH.  Finally, each PN data frame was corrected for air-mass 
and flux calibrated using standard FIGARO routines.  The flux calibration is 
not entirely rigorous for a variety of reasons, primarily because only one 
wide-slit exposure of the flux standard was obtained each night (i.e.\  one 
for each grating position), although the observing conditions were stable. 
Moreover, the flux calibration is based on an interpolated spectrum derived 
from calibrated data points spaced at intervals of less than one per observed 
wavelength range in a single order. Orders which contain prominent permitted 
transitions such as \ha\ and \hb, show wide absorption profiles in the 
standard star spectrum; these profiles may extend further than the limits of 
the observed wavelength range in that order and so may not have been properly 
corrected for.  Nevertheless, the internal consistency appears to be good, 
for example the total fluxes in the \ha\ and \hb\ lines, which appear in the 
spectra from both grating positions, were consistent to within 5\% and 
comparison of the total flux in the \niia\ emission line which appeared at 
grating position~1 with that in the \niib\ emission line observed at grating 
position~2 showed that the ratio of fluxes was within 6\% of the theoretical 
ratio of 3:1.  The absolute flux calibration was also found to be consistent 
with other observations.  An estimate of the total \hb\ flux from the nebula, 
obtained by considering the total flux observed through a single slit and 
scaling up from the slit area to the nebular area, is $8.4\times 10^{-14}$ J\ 
s$^{-1}$\ m$^{-2}$. This rather crude estimate compares well with the values 
in Acker \etal \shortcite{AckerCat} 
($9.33\times 10^{-14}$ J\ s$^{-1}$\ m$^{-2}$) and 
Harrington \etal \shortcite{HLWB97} 
($10.5\times 10^{-14}$ J\ s$^{-1}$\ m$^{-2}$). 
 
\section{star subtraction}
The central star continuum has been subtracted from the pv arrays displayed 
in this paper, so that the nebular emission can be appreciated.
This procedure is complicated by the fact that the central star of
\bd\ is a Wolf-Rayet type star with broad emission features
superimposed on the continuum spectrum.  The stellar emission features 
are about an order of magnitude broader than the nebular emission line 
features.  The two spatial pixels containing the brightest part of the
stellar spectrum were first extracted from the pv array
of the order of interest.  Any obvious nebular features were clipped
out of this spectrum and the resulting spectrum was fitted with a
polynomial function.  The fitted function was grown back into a 2-d
array.  A spatial profile of the central star was obtained  from the
original pv array (from a region well away from any nebular emission)
and this profile was used to modify the regenerated stellar pv array
in the spatial direction.  Finally this pv model for the stellar
spectrum was normalised to the same flux scale as the original pv
array by comparing the flux levels from the stellar continuum, again at a
position well away from any nebular features.  This pv array was then
subtracted from the original array to leave just nebular emission.
Careful analysis showed that small residuals appeared, corresonding to 
the extended 
wings of the stellar profiles at positions where broad stellar
emission features had occured, however, these were barely visible over 
the background noise level and we do not believe that they affect our
analysis of the nebular data.

\end{document}